# Correlations between Chondroitin Sulfate Physicochemical Properties and its in-vitro Absorption and Anti-inflammatory Activity

Lahari Surapaneni[1,2,*], G. Huang[3], Ashby B. Bodine[1], James R. Brooks[4], Ramakrishna Podila[5], and Vivian Haley-Zitlin[1]

1. Department of Food, Nutrition and Packaging Sciences, Clemson University, Clemson, SC 29634 USA
2. Department of Food Science, Pennsylvania State University, State College, PA 16803 USA
3. Department of Biosciences, Clemson University, Clemson, SC 29634 USA
4. Pharmavite, LLC. , Valencia, CA 91355 USA
5. Clemson nanomaterials center, Clemson, SC 29634 USA

*Corresponding author: lus29@psu.edu

**Abstract:** Here, we investigated the influence of physicochemical characteristics of chondroitin sulfate (CS) on its *in vitro* absorption and anti-inflammatory activity. We used eight different synthetic and natural CS samples with a range of molecular weights (7-35 kDa) and sulfation patterns. Our studies indicate that the absorption of CS is moderately correlated to percentage of chondroitin-6-sulfate while the anti-inflammatory activity may be weakly related to the molecular weight and the amount of total sulfation in the samples. Our *in vitro* studies could provide helpful screening tools for quick and effective evaluation of CS samples as a preliminary step towards *in vivo* studies.

**Introduction:** Osteoarthritis (OA) is a growing health concern in the United States. It is estimated that at least 67 million people will suffer from physician-diagnosed arthritis by 2030 (Issa & Sharma, 2006; Peat, McCarney, & Croft, 2001). Symptomatic therapies, employing non-steroidal anti-inflammatory drugs (NSAIDs), have been reported to be ineffective with questionable efficacy and adverse effects in many OA patients (Dieppe, 1978). The high price of other such pharmacological interventions as cyclooxygenase-2 (COX-2) inhibitors and local intra-articular steroid injections has outweighed their effectiveness or reduced side effects. In the recent past, nutritional supplements containing glucosamine and chondroitin sulfate (CS) have received much attention and support from researchers due to their promising potential in reducing pain and preserving joint functionality in patients with OA (McAlindon, LaValley, Gulin, & Felson, 2000). Numerous clinical studies provided evidence suggesting that orally administered CS acts as a slow acting symptomatic drug in patients with OA and is safe at doses as high as 800 – 1200 mg/day (Bucsi & Poor, 1998; Uebelhart, Thonar, Delmas, Chantraine, & Vignon, 1998). CS consists of a heterogeneous group of compounds with variable molecular weights and sulfation patterns (N. Volpi 2004). The animal-extracted CS samples, often found in many common supplements, are known to exhibit a variety of molecular weights and chondroitin-4-sulfate (CS4) to chondroitin-6-sulfate (CS6) ratio (N. Volpi, 2004). For instance, while CS samples extracted from porcine trachea exhibit a usual molecular weight ~18 kDa with a CS4-to-CS6 ratio of 2.4, shark cartilage derived CS samples possess >50 kDa molecular weight with a low CS4-to-CS6 ratio of 0.47. Despite the potential of CS in treating OA, very little is known regarding the influence of aforementioned heterogeneity and sulfation patterns on the bioavailability, anti-inflammatory activity, and in general overall efficacy of CS based supplements. Indeed, the heterogeneities in CS are crucial for some of their functions and have been found to strongly impact the role of CS in neuronal regeneration, plasticity, and recovery (V. P. Swarup et al. 2013). Therefore, a detailed study exploring the effects of heterogeneities in CS on its adsorption, anti-inflammatory, and anti-oxidative properties is necessary for identifying, designing, and developing new CS structures that could be biochemically synthesized. Although some speculations have been made regarding the correlation between molecular weight of CS and their bioavailability, at present, no such targeted studies have been conducted either *in vitro* or *in vivo*.

In this article, we investigated the influence of physicochemical characteristics of CS on its *in vitro* absorption and anti-inflammatory activity. We used eight different synthetic and natural CS samples with a range of molecular weights (7-35 kDa) and sulfation patterns. We employed Caco-2 monolayer cells to study the *in vitro* absorption and determine the permeability coefficients ($P_{eff}$) of different CS samples. Furthermore, the anti-inflammatory activity of CS samples was assessed by their ability to inhibit expression of a panel of inflammatory cytokines- tumor necrosis factor-alpha (TNF-α), and interleukin-6 (IL-6). Of the eight CS samples evaluated, four had a $P_{eff}$ value of 15 x $10^{-6}$ cm sec$^{-1}$ or higher indicating moderate to high absorption. Two of the four samples with higher $P_{eff}$ values were high molecular weight compounds (~ 35 kDa) suggesting possible influence of molecular weight on the bioavailability of CS. All the CS samples significantly inhibited expression of TNF-α (at both low ~5 mg/ml and high ~15 mg/ml concentrations) while the expression of IL-6 was inhibited by some of the CS samples at high concentration. Interestingly, we observed a moderate correlation between the absorption and percentage of 6-sulfated disaccharides in the CS samples while no correlation seemed to exist between the molecular weights and anti-inflammatory activity of these samples.

**Experimental:** We obtained 8 different CS samples labeled from A-F (4 synthetic, 3 bovine trachea-derives, 1 porcine trachea-derived) from Pharmavite, LLC, Northridge, CA (see Table 1). The physico-chemical characteristics of our samples including MW, CS4 and CS6 content were quantified using high performance size exclusion chromatography (HPSEC), enzymatic hydrolysis, and ion-pairing reversed-phase liquid chromatography.

*High performance size exclusion chromatography:* For HPSEC, two Agilent Bio SEC-3 (150 x 7.8 mm) columns with pore sizes 150 Å and 100 Å (Agilent, Part # 5190-2507 and # 5190-2502) were connected in tandem. 80% buffer (30 mM Disodium hydrogen phosphate, 30 mM Sodium sulfate, pH ~7) and 20% ethanol was used as mobile phase at a flow rate of 1.0 ml/min and run time of 15 minutes. 5 µl injection volumes of standards or samples were injected and chromatograms were recorded at 205 nm using a photo diode array (PDA) detector (Waters # 2998 PDA detector). A standard curve was plotted using retention times of standards and their corresponding molecular weights. The standard curve was then used to determine molecular weights of CS samples whose retention times were known (see Table 1).  Briefly, CS samples were incubated with chondroitinase AC II, an enzyme specific to chondroitin sulfate for 3 hours at 37 $^{o}$C to hydrolyze CS into its component disaccharides. Disaccharides were then separated and quantified by ion-pairing liquid chromatography with ultraviolet detection at 240 nm. While tetrabutylammonium bisulfate (340 mg; Sigma, Catalog # 86868) was dissolved in 1L of water (HPLC grade) for mobile phase A, 340 mg of tetrabutylammonium bisulfate was dissolved in 330 ml water and 670 ml acetonitrile (HPLC grade) for obtaining mobile phase B. Tris- (hydroxymethyl) aminomethane (TRIS) buffer solution was prepared by dissolving 3 g of TRIS (Sigma, Catalog # T1503), 2.4 g of anhydrous sodium acetate (Sigma, Catalog # S8750), 1.46 g of sodium chloride (ACS reagent grade), and 50 mg crystalline bovine serum albumin (Sigma, Catalog # A4378) in 100 ml of 0.12 M HCl. The pH of TRIS buffer was adjusted to 7.3 by adding adequate amount of 6M HCl (HPLC grade).

*Enzymatic hydrolysis:* For enzymantic hydrolysis, 5 units of chondroitinase AC II enzyme (Seikagaku America, Catalog # 100335-1A) were dissolved in 0.5 ml water and the stored at < 0 $^{o}$C when not in use. CS disaccharide reference standards (non-sulfated, CS4, and CS6 samples) were purchased from Sigma (Catalog # C3920, C4045 and C4170 respectively). The standard CS stock solutions were prepared by dissolving 2 mg of non-sulfated or 10 mg of CS4/CS6 each in 50 ml of water. A serial dilution of the stock solution was performed to obtain standard disaccharide solutions with concentrations of CS4 and CS6 between 2 - 100 µg/ml and non-sulfated samples between 0.4 - 20 µg/ml. Twenty µl of TRIS buffer solution, 30 µl of enzyme solution and 20 µl of CS sample were pipetted into 0.5 ml eppendorf tubes for enzymatic hydrolysis. The Eppendorf tubes were then incubated in a water bath at 37 $^{o}$C for 3 hours. Once removed from water bath, the samples in eppendorf tubes were allowed to cool to room temperature. 230 µl of mobile phase A was added to the tubes and the hydrolyzed CS solution was transferred into 2 ml HPLC vials. The eppendorf tubes were rinsed with another 200 µl of mobile phase A, which was transferred into the HPLC vial with hydrolyzed CS solution. Another 500 µl of mobile phase A was added to those vials and mixed well. Waters HPLC machine connected to a PDA detector (Waters, Part # 2996) was used to analyze the CS disaccharide standards and CS samples. Phenomenex

Synergi™ Polar-RP, 4.6 x 150 mm, 4 µm column was equilibrated with 80 % mobile phase A and 20% mobile phase B.

*In-vitro absorption studies*: We obtained ADMEcell CacoReady™ Caco-2 cell kits for *in vitro* intestinal absorption evaluation. This kit provided a 21-cell barrier in integrated HTS Transwell® -24 (Corning Lifesciences) with a proprietary shipping medium that is stable at room temperature. The kit of differentiated Caco-2 cell barriers was shipped on 13[th] day of differentiation and provided 14- day polarized cultures of Caco-2 cells on polycarbonate micro-porous filters in HTS Transwell® -24 plates (6.5 mm diameter, 0.33 $cm^2$ area and 0.4 µm pore diameter). For Caco-2 medium, 50 ml of fetal bovine serum (FBS) (Invitrogen, Catalog # 26140), 5 ml of 200 mM L-glutamine (Invitrogen, Catalog # 25030) and 5 ml of Penicillin/Streptomycin (Invitrogen, Catalog # 15150-148) were added to 440 ml of Dulbecco's modified eagle medium (DMEM) (Invitrogen, Catalog # 10567).Final concentrations of 30 mg/ml CS working samples were prepared by dissolving 150 mg of each CS sample in 50 ml sterile 1 X Hank's balanced salt solution (HBSS) transport buffer. HBSS transport buffer had a composition of 1 X Hank's balanced salt solution (HBSS) (Lonza, Catalog #b04-315Q) with 1.1 mM $MgCl_2 \cdot 6H_2O$ (Amresco, Catalog # E525) and 1.3 M $CaCl_2 \cdot 2H_2O$ (Amresco, Catalog # E506). Since CS is water soluble, all the CS samples dissolved completely in transport buffer. The samples were then aliquoted and stored at 2-8 $^{\circ}C$ until the day of permeability study. We used 0 µM – 20 µM propranolol hydrochloride solutions high permeability control and sodium Fluorescein (Sigma, Catalog # 46960) as a low permeability control. A 200 µM Lucifer yellow CH dilithium salt (Sigma, Catalog # L0259) solution was prepared and was used to test integrity of the Caco-2 cell monolayer.

On the day of permeability experiments, transepithelial electrical resistance (TEER) measurement was performed using an Epithelial Volt-Ohm meter (EVOM) (World precision instruments, Catalog # EVOM2) on the CacoReady™ plate prior to any further processing. Any monolayers with TEER > 1000 Ohm. $cm^2$ were considered acceptable for permeability assays. The CS samples, blank transport buffer, high permeability and low permeability controls, and Lucifer yellow were all warmed to 37 $^{\circ}C$ prior to the experiments. The basal compartment of CacoReady™ plate was washed with 750 ul of transfer buffer per well. Following aspiration of the transport buffer wash, 750 µl of transport buffer was aliquoted per well again. Cell culture medium from apical compartment wells was aspirated, discarded, and the apical wells were gently washed using 300 µl of transport buffer per well. Subsequently, 300 µl of CS samples, controls and blank transport buffer were filled in the apical wells. CacoReady™ plate with samples was placed in the incubator at 37 °C, 5% $CO_2$ for 2 hours. After 2 hrs, samples from apical wells and basal wells of the CacoReady™ plate were then collected separately into 1 ml Eppendorf tubes. Quantitative analysis of CS in apical and basal samples was done by enzyme hydrolysis and liquid chromatography. Apical and basal concentrations of propranolol, sodium fluorescein and lucifer yellow were analyzed by Fluorescence spectrophotometry (SpectraMax M2e, Molecular Devices, LLC., Sunnyvale, CA).

*Inflammatory response*: The influence of CS on the production of inflammatory markers such as TNF-α (Invitrogen, Catalog # 3012), IL-1β, IL-6, and NO production (Griess reagent) was measured on RAW264.7 cells using standard ELISA and Griess reagent kits.

**Results and Discussion**: As shown in Table 1, the molecular weights of sample A-F, obtained using HPSEC, were found to range from 7-35 KDa. Considering the limited sensitivity of the HPSEC in separating samples very close to each other in size, it was difficult to obtain the absolute molecular weights of these samples. Nonetheless, it could be gleaned from Table 1 that the sample pairs A-B, C-D, E-F, and G-H possess similar molecular weights. Although the samples differed from each other in % ΔDi-6S, they do not differ much in total sulfation (= % ΔDi-4s+% ΔDi-6S). Importantly, samples B, G and H had higher % ΔDi-6S compared to the other CS samples used in this study.

Considering its morphological and functional similarities to the enterocytes of small intestine, Caco-2 has been used as a well-established in vitro method for evaluating drug/nutrient absorption. Although the bioavailability of a drug is hard to be quantified using Caco-2 studies, it is still helpful to qualitatively rank a series of compounds (in this case, samples A-F) in order of their possible absorbability.

Any compounds with $P_{eff}$ values over $10^{-6}$ cm s$^{-1}$ are considered to be of high absorption with over 90% absorption while compounds with $P_{eff} < 10^{-7}$ cm s$^{-1}$ have a very low absorption (Artursson et al., 2001).

As shown in Fig. 1a, Sample B exhibited the highest $P_{eff}$ among the samples. Although samples A and B were of similar molecular weight, their $P_{eff}$ values were found to different with $P_{eff}$ for B > $P_{eff}$ for A. Such an observation may possibly be attributed to the fact that % ΔDi- 6S in sample B is higher to sample A. Despite of their comparatively higher molecular weights, samples G and H showed higher absorption than the lower molecular weight samples A, C, D and E and whose % ΔDi- 6S was lower than that of samples G and H. Fig. 1b showed that no correlation exists between molecular weight and $P_{eff}$ suggesting that the absorption of CS is independent of molecular weight. The bivariate fit of $P_{eff}$ and % ΔDi- 6S in CS sample resulted in *R*-square value of 0.5 (weak correlation) and a p-value ~0.0477 revealing that CS absorption may depended upon % ΔDi- 6S and alternatively the source material of CS. (Fig. 1c). Such an observation may possibly explain the earlier results wherein shark chondroitin sulfate exhibited a higher absorption despite its high molecular weight of 44 kDa seen in humans (N. Volpi, 2003). A lack of natural or synthetic CS samples with pure ΔDi-4S in our sample set, no correlation data could be obtained for $P_{eff}$ and % ΔDi-4S. Nonetheless, the significant effect of % ΔDi-6S on $P_{eff}$ (Fig. 1c) combined with a lack of correlation between % total sulfated ΔDi and $P_{eff}$ (Fig. 1d) indicates that the presence of non-zero amounts of 4S may negate positive correlation of ΔDi-6S with $P_{eff}$ in CS.

We observed that all CS samples exhibited moderate to strong anti-inflammatory activity by inhibiting inflammatory cytokine (TNF-α and IL-6) production by the RAW 264.7 cells. Our results are in agreement with cytokine lowering effect of CS as reported by several researchers in *in vitro* studies (G. M. Campo et al., 2009), animal (Bauerova et al., 2011; Giuseppe M. Campo et al., 2003) studies. As shown in Table 2 and Fig. 2, the % reduction in TNF-α may be very weakly correlated to MW and % ΔDi-total since it exhibited R2 values > 0.42. Furthermore, it is possible that % reduction in TNF-α is related to % ΔDi-4S since no (weak) correlation existed between TNF-α and % ΔDi-6S (% ΔDi-total). Interestingly, the correlation between TNF-α and MW/% ΔDi-total were found to be more pronounced at higher CS concentrations (15 µg/ml). We did not find any systematic correlations had been observed between the molecular weights, % ΔDi-6S or total % sulfated ΔDi and IL-6 levels. At 5 µg/ml concentration, sample B ranked high among the samples in inhibiting the production of LPS stimulated TNF-α in RAW 264.7 cells. However, at the same concentration, the inhibition of IL-6 by sample B was not significant compared to LPS positive. At 15 µg/ml concentration, sample B stands out as the best among the CS samples by inhibiting production of both TNF-α and IL-6. These studies suggest the anti-infammatory activity of CS, unlike its absorption, does not systematically depend on any physicochemical parameters.

**Conclusions:** Synthetic and natural CS samples with a range of molecular weights (7-35 kDa) and sulfation patterns showed that the absorption of CS may be weakly correlated to % ΔDi-6S and is nearly independent of molecular weight. All the CS samples significantly inhibited expression of TNF-α (at both low ~5 mg/ml and high ~15 mg/ml concentrations) while the expression of IL-6 was inhibited by some of the CS samples at high concentration. We observed a weak correlation between the % reduction in TNF-α and MW and % ΔDi-total while no correlation seemed to exist between the IL-6 and physicochemical characteristics of CS. In conclusion, our studies provide a guiding tool in identifying the CS compounds that are most likely to be effective *in vivo* or in clinical studies. The *in vitro* screening strategies described in this article could help lower the expense incurred on the clinical trials by several folds by helping in sample pre-screening in terms of absorption and anti-inflammatory activity.

Figures and Figure Captions:

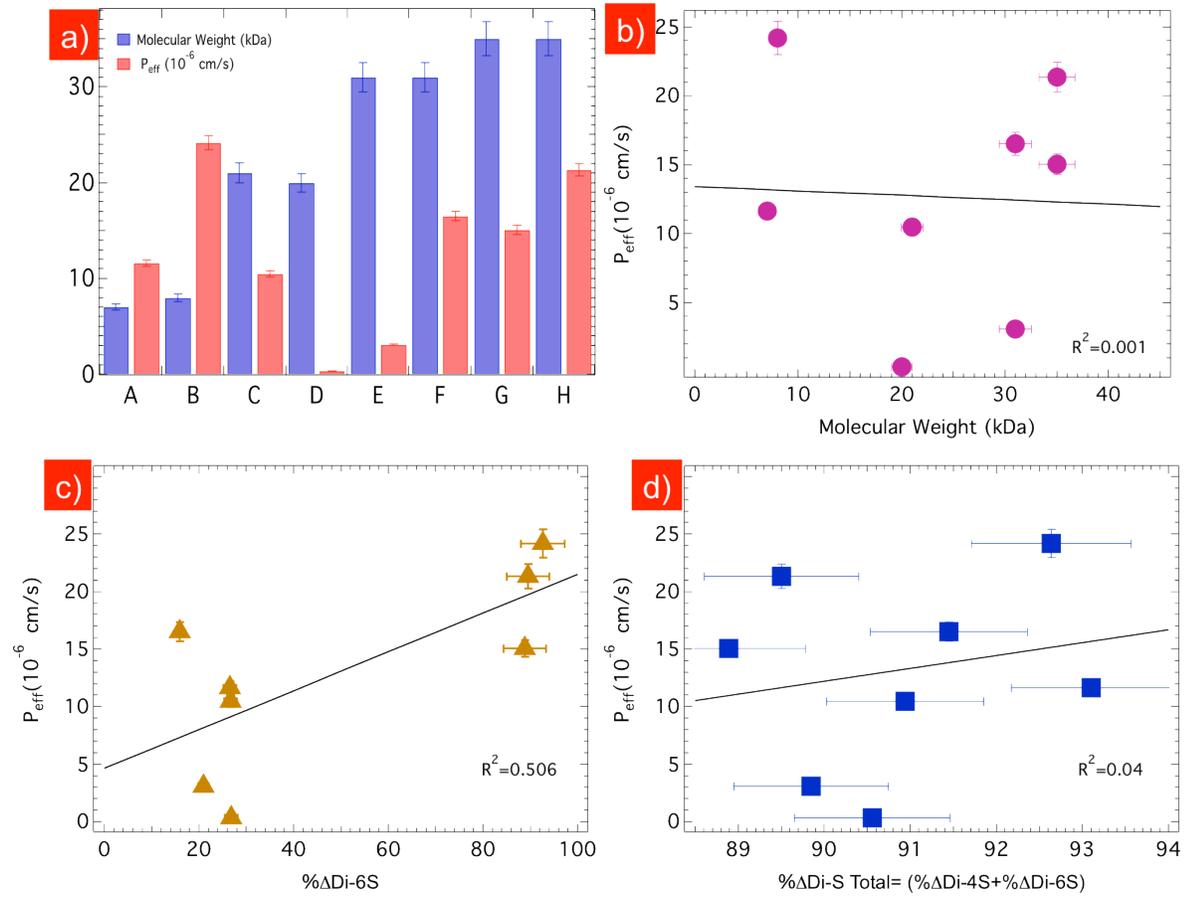

Figure 1: The trends for Caco-2 cell effective permeability ($P_{eff}$) as a function of molecular weight (a and b), %6S (c), and total sulfation (d).

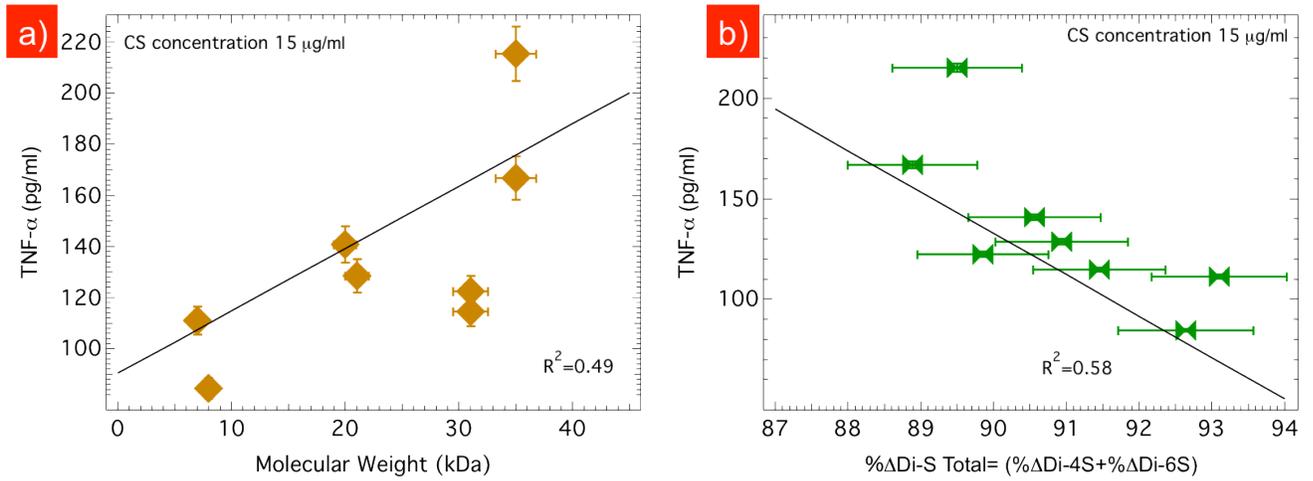

Figure 2: The trends for moderately correlated TNF-a (MW and %total sulfation) from RAW 264.7 macrophages. The correlation coefficients for TNF-a on all physicochemical parameters are shown in Table 2.

| Sample ID | MW (kDa) | %ΔDi-0s | %ΔDi-4s | %ΔDi-6s |
|---|---|---|---|---|
| A | 7 | 6.91 | 66.61 | 26.49 |
| B | 8 | 7.36 | 0 | 92.64 |
| C | 21 | 9.06 | 64.21 | 26.73 |
| D | 20 | 9.44 | 63.73 | 26.83 |
| E | 31 | 10.14 | 68.84 | 21.01 |
| F | 31 | 8.56 | 75.58 | 15.87 |
| G | 35 | 11.11 | 0 | 88.89 |
| H | 35 | 10.56 | 0 | 89.5 |

Table 1: Molecular weight and the content of unsulfated, chondroitin-4-sulfate, and chondroitin-6-sulfate, determined using HPSEC and HPLC measurements.

|         | MW (kDa) |      | %ΔDi-6s |      | %ΔDi-Total |      |
|---------|----------|------|---------|------|------------|------|
| CS conc. (µg/ml) | 5 | 15 | 5 | 15 | 5 | 15 |
| TNF-α   | 0.42     | **0.49** | 0.06 | 0.15 | 0.44 | **0.58** |
| IL-6    | 0.11     | **0.52** | **0.49** | 0.22 | 0.34 | 0.42 |

Table 2: Pearson correlation coefficients $R^2$ for % reduction in TNF-a, IL-6 and various physicochemical properties at two different concentrations of CS.